\documentclass[twocolumn,prl,preprintnumbers,tightenlines,footinbib,superscriptaddress]{revtex4}

\pdfoutput=1
\usepackage{amsmath,amssymb,amsfonts}
\usepackage{graphicx}
\usepackage[normalem]{ulem}
\usepackage{slashed}
\usepackage{booktabs}
\usepackage{hyperref}
\usepackage{color}
\usepackage[sort&compress]{natbib}
\usepackage{tikz}
\usepackage{cleveref}
\newcounter{qnumber}

\newcommand{\drawsquare}[2]{\hbox{%
\rule{#2pt}{#1pt}\hskip-#2pt
\rule{#1pt}{#2pt}\hskip-#1pt
\rule[#1pt]{#1pt}{#2pt}}\rule[#1pt]{#2pt}{#2pt}\hskip-#2pt
\rule{#2pt}{#1pt}}

\newcommand{\Yfund}{\raisebox{-.5pt}{\drawsquare{6.5}{0.4}}}
\newcommand{\Ysymm}{\raisebox{-.5pt}{\drawsquare{6.5}{0.4}}\hskip-0.4pt%
        \raisebox{-.5pt}{\drawsquare{6.5}{0.4}}}
\newcommand{\Yasymm}{\raisebox{-3.5pt}{\drawsquare{6.5}{0.4}}\hskip-6.9pt%
        \raisebox{3pt}{\drawsquare{6.5}{0.4}}}

\begin{document}

\title{The Phases of Non-supersymmetric Gauge Theories: the $SO(N_c)$ Case Study}

\author{Csaba Cs\'aki}
\email{ccsaki@gmail.com}
\affiliation{Department of Physics, LEPP, Cornell University, Ithaca, NY 14853, USA}

\author{Andrew Gomes}
\email{awg76@cornell.edu}
\affiliation{Department of Physics, LEPP, Cornell University, Ithaca, NY 14853, USA}

\author{Hitoshi Murayama}
\email{hitoshi@berkeley.edu, hitoshi.murayama@ipmu.jp, Hamamatsu Professor}
\affiliation{Department of Physics, University of California, Berkeley, CA 94720, USA}
\affiliation{Kavli Institute for the Physics and Mathematics of the
  Universe (WPI), University of Tokyo,
  Kashiwa 277-8583, Japan}
\affiliation{Ernest Orlando Lawrence Berkeley National Laboratory, Berkeley, CA 94720, USA}

\author{Ofri Telem}
\email{t10ofrit@gmail.com}
\affiliation{Department of Physics, University of California, Berkeley, CA 94720, USA}
\affiliation{Ernest Orlando Lawrence Berkeley National Laboratory, Berkeley, CA 94720, USA}

\begin{abstract} We investigate the IR phases of non-supersymmetric (non-SUSY) $SO(N_c)$ gauge theories with $N_F$ fermions in the vector representation obtained by perturbing the SUSY theory with anomaly mediated SUSY breaking (AMSB). We find that of the wide variety of phases appearing in the SUSY theory only two survive: for $N_F\leq \frac{3}{2} (N_c-2)$ the theory confines, breaking the $SU(N_F)$ global symmetry to $SO(N_F)$, while for $\frac{3}{2} (N_c-2)<N_F<3(N_c-2)$ the theory flows to a (super)-conformal fixed point. The abelian Coulomb and free magnetic phases do not survive and collapse to the confining phase. We also investigate the behavior of loop operators in order to provide a clear distinction between the confining and screened phases. With the choice of $Spin(N_c)$ for the global structure of the gauge group, we find that the electric Wilson loop indeed obeys an area law, providing one of the first demonstrations of true confinement with chiral symmetry breaking in a non-SUSY theory. We identify monopole condensation as the dynamics underlying confinement. These monopoles arise naturally for $N_F=N_c-2$. The case with smaller number of flavors can be obtained by integrating out flavors, and we confirm numerically that the monopole condensate persists in the presence of AMSB and mass perturbations. 
\end{abstract}

\maketitle

\section{Introduction}
Understanding the phases of strongly coupled gauge theories is one of the most important outstanding goals of particle physics. A great deal of progress was made on this problem in the mid-90's, led by Seiberg~\cite{Affleck:1983mk,Seiberg:1994bz,Seiberg:1994pq}, when the vacuum structures of a large variety of supersymmetric (SUSY) gauge theories were established, including ${\cal N}=1$ SUSY QCD, ${\cal N}=2$ Seiberg-Witten theories, and also various interesting chiral gauge theories. A wide variety of gauge theory phases were encountered in this process: the confining phase, the screened/Higgs phase (which are indistinguishable due to complementarity), a free magnetic phase with an IR free dual gauge group, true conformal field theories (CFTs) with a non-trivial interacting fixed point dubbed the non-abelian Coulomb phase, and finally the abelian Coulomb phase, which in the SUSY case usually includes points in the moduli space with massless monopoles \cite{Seiberg:1994rs}. These phases form an intricate web which can be nicely connected by integrating out flavors or Higgsing the theories. One particularly interesting example is the case of $SO(N_c)$ gauge theories with $N_f$ chiral superfields in the vector $N_c$ representation: as $N_f$ is varied between $1$ and $3N_c-2$ all of the above phases actually appear, making $SO(N_c)$ an ideal test lab for studying the phases of gauge theories.  

SUSY gauge theories are very interesting in their own right, though one would obviously like to ask what phases appear for the non-SUSY versions of these theories. This question was investigated in a series of papers in the late 90's \cite{Evans:1995ia,Aharony:1995zh,Evans:1995rv,DHoker:1996xdz,Alvarez-Gaume:1996vlf,Alvarez-Gaume:1996qoj,Evans:1996hi,Konishi:1996iz,Alvarez-Gaume:1997bzm,Evans:1997dz,Alvarez-Gaume:1997wnu,Cheng:1998xg,Martin:1998yr,Luty:1999qc}, by perturbing SUSY gauge theories with soft supersymmetry breaking terms (see also the more recent \cite{Cordova:2018acb}). In~\cite{Murayama:2021xfj}, a simple and efficient method was introduced for studying non-supersymetric gauge theory by starting with its supersymmetric version and perturbing it by anomaly mediated SUSY breaking (AMSB) \cite{Randall:1998uk,Giudice:1998xp} (see also \cite{ArkaniHamed:1998kj,Arkani-Hamed:1998dti} for earlier work containing some important aspects of AMSB). The main advantage of this method is that a single spurion provides both the positive scalar mass squared and the gaugino mass, thereby preparing exactly the UV theory of interest. By the UV insensitive nature of AMSB the determination of the low-energy phase of the theory can be reliably performed within the dual low-energy effective theory. Indeed many interesting results have been obtained this way including the phases and global symmetry breaking pattern of SUSY QCD~\cite{Murayama:2021xfj} as well as in the simplest non-supersymmetric chiral gauge theories with an antisymmetric~\cite{Csaki:2021aqv} or symmetric~\cite{Csaki:2021xhi} fermion. These results allowed the reexamination of old conjectured phases of non-SUSY chiral gauge theories based on the tumbling/most attractive channel (MAC) framework~\cite{Raby:1979my,Dimopoulos:1980hn}. It was found that all cases needed some form of modification, and in some cases it was not the condensate in the MAC that was actually generated. The theories with larger number of flavors included more chiral symmetry breaking and also examples of superconformal fixed points from non-SUSY UV theories. 

However all of the theories explored so far with AMSB contained some particles in the fundamental representation, which led to screening and a perimeter law for all of their Wilson loops. Consequently, their chiral symmetry breaking took place in a screened/Higgs phase, rather than a genuine confining phase. This will no longer be the case for gauge theories with Lie algebra $\mathfrak{so}(N_c)$ (more specifically $Spin(N_c)$) and matter in the vector representation, where the Wilson loop in the spinor can not be screened. This makes the appearance of an area law possible -- which is the strict definition of a confining phase.

In this paper we study the low-energy dynamics of the $SO(N_c)$ gauge theories with $N_F$ Weyl fermions in the vector representation, obtained by perturbing the corresponding ${\cal N}=1$ $SO(N_c)$ theory via AMSB. We will also include a discussion of the global properties of these theories, which is necessary to distinguish the truly confining theories with an area law for the Wilson loop from those with a perimeter law. Thus, as will be explained later, we will have to distinguish between $Spin(N_c)$ and the $SO(N_c)_\pm$ groups. The $N_F=N_c-2$ case will be particularly interesting: the SUSY theory in this case corresponds to a pure Coulomb branch, where a pair of monopoles becomes massless at one point, together with a full multiplet of dyons becoming massless at the origin. We will show that the AMSB perturbation of this theory leads to the condensation of monopoles, along with (partial) breaking of the chiral global symmetries at the true ground state. Such monopole (or dyon) condensation was conjectured by Mandelstam \cite{Mandelstam1976} and 't Hooft \cite{tHooft:1981bkw} long ago to be the dynamics leading to confinement via the dual Meissner effect. This has indeed been found to be the case by Seiberg and Witten when perturbing pure ${\cal N}=2$ to pure ${\cal N}=1$ supersymmetric Yang--Mills (SYM)~\cite{Seiberg:1994rs}, and also by~\cite{Aharony:1995zh,Alvarez-Gaume:1996vlf,Konishi:1996iz,Evans:1995rv} who studied the small non-SUSY perturbations of the Seiberg-Witten theory. Our results provide one of the first examples of (true) confinement with chiral symmetry breaking in a non-SUSY gauge theory.

By considering mass deformations to the $N_F=N_c-2$ case, we will be able to show that all of the cases with $N_F< \frac{3}{2}(N_c-2)$ exhibit the same behavior as for $N_F=N_c-2$, {\it i.e.}\/, electric confinement via monopole condensation, together with chiral symmetry breaking. In contrast, for $\frac{3}{2}(N_c-2)<N_F\leq3(N_c-2)$ the non-SUSY UV theory flows as expected to a (super)-conformal theory in the IR. The case with  $N_F=\frac{3}{2}(N_c-2)$ is marginal, and will be discussed in future work. In other words, with the AMSB perturbation the many phases present in the SUSY case collapse down to just two: the confining phase, and the conformal phase. The Coulomb and free magnetic phases do not survive the breaking of supersymmetry. 

The paper is organized as follows. First we present a short summary of the moduli space and symmetries of ${\cal N}=1$ SUSY $SO(N_c)$ theories with $N_F$ vectors. The quantum vacuum structure of the entire SUSY series has been worked out in a beautiful paper by Intriligator and Seiberg in~\cite{Intriligator:1995id}, which will be the basis of our analysis for the AMSB perturbations. We begin our discussion of the various cases with the most novel $N_F=N_c-2$ case, which is the only one known example so far giving rise to monopole condensation with chiral symmetry breaking in a non-SUSY theory. We then show that whenever ``non-trivial" (spinorial) Wilson lines exist in the theory with $N_F=N_c-2$, they exhibit an area law, signaling true electric confinement. By considering mass deformations to the $N_F=N_c-2$ case, we demonstrate this fact, as well as chiral symmetry breaking, for all $SO(N_c)$ theories with $1<N_F\leq N_c-2$. While there are several special cases to examine following the analysis of~\cite{Intriligator:1995id}, in the end the global minimum of the theory for $1<N_F\leq N_c-2$ is always the one with the $SU(N_F) \rightarrow SO(N_F)$ chiral symmetry breaking pattern. We have also summarized this important result in a companion letter \cite{Csaki:2021jax}.

We then continue on to consider the cases with a larger number of flavors. We find results similar to the case of SUSY QCD. For $N_c-2< N_F \leq \frac{3}{2} (N_c-2)$ we again find a global minimum where the chiral symmetry is broken to $SO(N_F)$. In this case monopole condensation does not directly appear in the description, but non-trivial Wilson loops still exhibit an area law. To see this, note that at a generic point in the moduli space of the dual theory, the dual quarks are integrated out, leaving us with pure $SO(N_F-N_c+4)$ SYM. Consequently, whenever the dual gauge group allows for dyonic ('t Hooft) loops, these exhibit an area (perimeter) law. By the duality of \cite{Aharony:2013}, whenever the original theory allows for non-trivial Wilson ('t Hooft) loops, they exhibit an area (perimeter) law.

Finally, for $\frac{3}{2}(N_c-2)\leq N_F \leq 3 (N_c-2)$ we find that the theory flows to a superconformal fixed point, even though the UV theory explicitly breaks supersymmetry. For $3 (N_c-2)<N_F$, the theory with AMSB has tachyonic squarks and hence no ground state, and is not continuously connected to the non-supersymmetric $SO(N_c)$ theory.

\section{$SO(N_c)$ with $N_F$ Fundamentals}
We consider a supersymmetric gauge theory with gauge group $SO(N_c)$, $N_F$ vectors, and no tree-level superpotential. This theory has been thoroughly studied in {\it e.g.}\/, \cite{Intriligator:1995id,Intriligator:1995au}. The anomaly-free global symmetry of the theory is $\left(SU(N_F)\times U(1)_R\times \mathbb{Z}_{2N_F}\times \mathbb{Z}_2/\mathbb{Z}_{N_F}\right)$, where the $\mathbb{Z}_2$ is charge conjugation \footnote{For $N_F=3$ the discrete part of the global symmetry is enhanced from $Z_{2N_F}$ to $Z_{4N_F}$}.  Under the continuous part of the global symmetry, the matter fields transform as $Q(\Yfund)_{\frac{N_F-N_c+2}{N_F}}$.

For $N_F<N_c$, the $D$-flat directions of the theory are given, up to gauge and global transformations, by
\begin{align}
	&Q = \left( \begin{array}{c|c}
		\begin{array}{ccc}\varphi_1&&\\&\ddots&\\&&\varphi_{N_F}\end{array}&0
		\end{array} \right)\, ,
\end{align}
For $N_F<N_c$ $D$-flat directions are conveniently parameterized by the $\frac{1}{2}N_F(N_F+1)$ gauge invariant ``meson" operators $M^{ij}=Q^{i}Q^{j}$.  Along the $D$-flat directions, they are given by
\begin{align}
	&M = \text{diag}\left(\varphi^2_1,\ldots,\varphi^2_{N_f}\right)\,.
\end{align}
For  $N_F\geq N_c$, the $D$-flat direction is given by
\begin{align}
	&Q = \left( \begin{array}{c}
		\begin{array}{ccc}\varphi_1&&\\&\ddots&\\&&\varphi_{N_c}\end{array}\\\hline0
		\end{array} \right)\, 
\end{align}
It is conveniently parameterized in terms of meson operators, as well as the baryon operators $B^{[i_1,\ldots i_{N_c}]}=Q^{[i_1}\ldots Q^{i_{N_c}]}$, where the $i$ are flavor indices and we take the gauge singlet out of the tensor product of $N_c$ fundamentals of $SO(N_c)$. The gauge invariant operators are given, up to global transformations, by
\begin{align}
	&M = \left( \begin{array}{c|c}
		\text{diag}\left(\varphi^2_1,\ldots,\varphi^2_{N_c}\right)&0\\\hline
		0&0_{{N_F-N_c}\times {N_F-N_c}}
		\end{array} \right)\nonumber\\
	&B^{1,\ldots {N_c}} = \varphi_1\ldots \varphi_{N_c}\,.
\end{align}
The IR behavior of the theory strongly depends on the relative magnitudes of $N_c$ and $N_F$ and is summarized in Table~\ref{tab:summary}. Below we will show that adding AMSB to the theory leads to chiral symmetry breaking for all $1< N_F\leq \frac{3}{2}(N_c-2)$. Furthermore, in this range the theory confines; below we give an exact meaning to this statement in terms of the loop operators of the theory. We assume throughout that $N_c>3$, and leave the $N_c=3$ to future work.

\begin{table}[t]
	\centerline{
	\begin{tabular}{|c|c|c|} \hline
	Range& SUSY &+AMSB \\ \hline
	$N_F=1$ & run-away  & confinement \\ \hline
	$1<N_F<N_c-4$ & run-away &  confinement$+ \chi$SB \\ \hline
	$N_F=N_c-4$ & 2 branches & confinement$+ \chi$SB \\ \hline
	$N_F=N_c-3$ & 2 branches & confinement$+ \chi$SB\\ \hline
	$N_F=N_c-2$ & Coulomb & confinement$+ \chi$SB \\ \hline
	$N_F=N_c-1$ & $\begin{array}{c}\text{free magnetic}\\\text{2 branches} \end{array}$& confinement$+ \chi$SB\\ \hline
	$N_F=N_c$ & $\begin{array}{c}\text{free magnetic}\\\text{2 branches} \end{array}$ & confinement$+ \chi$SB\\ \hline
	$N_c+1\leq N_F\leq \frac{3}{2}(N_c-2)$& free magnetic &  confinement $+ \chi$SB\\ \hline
	$\,\frac{3}{2}(N_c-2)<N_F\leq 3(N_c-2)\,$ & CFT & CFT \\ \hline
	$3(N_c-2)<N_F$& IR free & run-away \\ \hline
	\end{tabular}
	}
	\caption{Summary of IR Behavior of $SO(N_c)$ theories with $N_F$ fundamentals with AMSB. $\chi$SB stands for chiral symmetry breaking. For $N_F=N_c-1$ and $N_c$, two branches appear along the flat direction of the maximum rank of the meson $M^{ij}$, yet the AMSB chooses one over the other, resulting in the $\chi$SB.}\label{tab:summary}
\end{table}

\section{Phases of Gauge Theories}

One often hears the word ``confinement" describing the situation in which colored degrees of freedom are bound into color-singlet states, even if a linear potential is lacking due to screening from quark-antiquark production. We will be more careful with the word and, following \cite{Aharony:2013}, only use it in the narrow context of a particular Wilson/'t Hooft/dyonic loop operator \footnote{By a loop operator we mean a closed line around which we take a test charge, monopole or dyon in some representation allowed by the gauge group.} exhibiting an area law, in which case we will say that the given loop operator confines.  We are especially interested in the confinement of non-trivial loop operators - the ones which transform non-trivially under the center of the gauge group. Note that in some of the literature, {\it e.g.}\/, \cite{Aharony:2013}, these closed loops are referred to as line operators - we will use the more conventional name ``loop operators" to stress their gauge invariance.

The allowed non-trivial loop operators in the theory depend on the choice of the global properties of the gauge group -- for example, with the Lie algebra $\mathfrak{so}(N_c)$ the gauge group can be $Spin(N_c),\,SO(N_c)$, and so on. Depending on the choice of gauge group, the allowed non-trivial loop operators can be Wilson, 't Hooft, or dyonic loops. Whatever the choice may be, these loops exhibit either a perimeter or an area law, depending on the local physics, which is in and of itself insensitive to the global properties of the gauge group.

One possible choice of the gauge group is $Spin(N_c)$, which is the universal cover of all Lie groups that share the Lie algebra $\mathfrak{so}(N_c)$. In this case the non-trivial loop operators are Wilson loops, while others are forbidden by Dirac quantization. An area law for non-trivial Wilson loops indicates the confinement of the electric degrees of freedom associated with it.
Other choices for the global structure are obtained by modding out the $Spin(N_c)$ by subgroups of its center, which is $\mathbb{Z}_2$ for odd $N_c$ and larger for even $N_c$ \cite{Aharony:2013}. Here we only consider $SO(N_c)=Spin(N_c)/\mathbb{Z}_2$. In this case the non-trivial loop operators are either purely magnetic 't Hooft loops, a choice denoted by $SO(N_c)_+$, or dyonic loops, a choice denoted by $SO(N_c)_-$. In each case, other non-trivial loop operators are forbidden by Dirac quantization. Below, when we speak of ``the loop" in a particular theory, we will be referring to the single non-trivial loop that the theory admits, whether electric, magnetic, or dyonic. Additionally, when we are only concerned with the local physics we will simply refer to the gauge group as $SO(N_c)$.

While $Spin(N_c)$ and $SO(N_c)_\pm$ exist on equal footing as possible gauge groups, we will be particularly interested in $Spin(N_c)$, as it can provide what eludes us for $SU(N)$ gauge theories with fundamental matter: an order parameter for electric confinement. Whereas the fundamental Wilson loop of $SU(N)$ (which is its own universal cover) can be screened by the fundamental matter, the spinorial Wilson loop of $Spin(N_c)$ cannot be screened by the vectorial quarks. 

The main objective of our study is to determine the phases of supersymmetric $SO(N_c)$ gauge theories when perturbed by AMSB. We will be able to determine the local behavior of the theory -- be it chiral symmetry breaking, monopole condensation, etc. As a result, we will be able to establish the behavior of the allowed non-trivial loop operators in the theory, and whether they exhibit an area or a perimeter law. Our final results about the phase structure (along with the corresponding SUSY phases) are summarized in Tab.~\ref{tab:summary}. We can see that in the non-SUSY theory the only surviving phases are those of confinement (with chiral symmetry breaking), or a conformal phase. The abelian Coulomb and the free magnetic phases do not survive the AMSB perturbation, and they all collapse to the generic confining+$\chi$SB phase. 

\section{Confinement with Chiral Symmetry Breaking for $1< N_F\leq \frac{3}{2}(N_c-2)$}

Next we present a detailed analysis of the vacuum structure of $SO(N_c)$ with $N_F$ flavors in the presence of AMSB. For $N_F=N_c-2$ we show monopole condensation with chiral symmetry breaking -- leading to the confinement of non-trivial Wilson lines for $Spin(N_c)$\footnote{Alternatively, for $SO(N_c)_+$ the non-trivial 't Hooft line exhibit a perimeter law, while for $SO(N_c)_-$, the non-trivial dyonic line exhibits an area law.}. For $N_F<N_c-2$ we can still explicitly see chiral symmetry breaking, while monopole condensation and thus electric confinement is established by adding mass deformations to the $N_F=N_c-2$ case. Finally, theories with $N_c-2 < N_F\leq \frac{3}{2}(N_c-2)$ exhibit chiral symmetry breaking, while confinement is demonstrated by finding a ``hidden" monopole condensate in the dual theory.

\subsection{$N_F=N_c-2$}
In this case the supersymmetric theory is in an abelian Coulomb phase \cite{Intriligator:1995id}. Since the $M^{ij}$ are not charged under $U(1)_R$, there is no superpotential even at the quantum level, and hence the theory has a quantum moduli space. On this moduli space, the gauge symmetry is higgsed to a $SO(2) \simeq U(1)$, namely, the theory is on the Coulomb branch. On the moduli space, the gauge coupling $\tau=\frac{\theta}{2\pi}+\frac{i8\pi}{g^2}$  is given only as a function of the $SU(N_F)$ invariant $U\equiv\text{det}M$. It is singular at two points $U=0$ ($U=U_1\equiv16\Lambda^{2N_F}$), where the dyons $q^{\pm}_i$ (monopoles $E^{\pm}$) of the $U(1)$ gauge symmetry become massless. In the original paper \cite{Intriligator:1995id}, the authors chose to label the particles condensing at $U=0$ and $U=U_1$ as monopoles and dyons, respectively. Our opposite labeling leads to line behaviors consistent with those in \cite{Aharony:2013} for all confining theories, and is also consistent with the finding in \cite{Witten:2000nv}. 

Around the singular point $U=0$ the relevant light degrees of freedom are the dyons $q^{\pm}_i$ with magnetic charge $\pm1$, which transform under the UV global symmetry $SU(N_F)\times U(1)_R$ as $q^{\pm}_i(\overline{\Yfund})_1$.
These have a dynamically generated superpotential about $U=0$ of
\begin{eqnarray}\label{eq:monYuk}
W_{\text{dyon}}\,&=&\,\frac{1}{\mu}\,f(t)\,M^{ij}q^+_{i}q^-_{j}\,,
\end{eqnarray}
where $\mu$ is an effective mass scale, $t=U\Lambda^{4-2N_c}$, and $f(t)$ is a holomorphic function in the neighborhood of $t=0$, normalized so that $f(0)=1$. Expanding $f$ to higher orders in $t$ introduces tree level AMSB, but it is highly suppressed by powers of the meson VEV over $\Lambda$ and results in no qualitative changes. Exactly at $U=0$, 't Hooft anomaly matching is saturated by $q^{\pm}_i,\,M^{ij}$, and the photinos $\mathcal{W}_\alpha\sim W_\alpha Q^{N_c-2}$ \cite{Intriligator:1995id}, whose charges are given in Table~\ref{tab:Ncm2U0}.

\begin{table}[t]
		\centerline{
		\begin{tabular}{|c|c|c|c|} \hline
	& $SO(N_c)$ & $SU(N_F)$&$U(1)_R$ \\ \hline
	$Q^i$ & $\Yfund$ & $\Yfund$&$0$\\ \hline 
	$\lambda$&$\Yasymm$&$\bf 1$&$1$\\ \hline\hline
	 $M_{ij}$&$\bf1$&$\Ysymm$&$0$\\ \hline
	 $q^{\pm}_i$&$-$&$\overline{\Yfund}$&$1$\\ \hline
	 $\lambda_{\text{mag}}$&$-$&$\bf 1$&$1$\\ \hline
	\end{tabular}
	}
	\caption{DOF of the $SO(N_c)$ theory with $N_F=N_c-2$ near $M=0$. $\lambda$ are the $SO(N_c)$ gauginos, while $\lambda_{\text{mag}}$ are the photinos of the unbroken (magnetic) $U(1)$ in the IR. For the supersymmetric theory at the origin, the full global symmetry is unbroken. With AMSB there is a local minimum, where the global symmetry is broken to $SU(N_F-2)$. }\label{tab:Ncm2U0}
\end{table}

Using the formulae for loop level AMSB Eq.~(\ref{eq:AMSBloop}), we can explore the local minima around the origin of moduli space. The IR free nature of the $U(1)$ gauge theory gives a tachyonic contribution to the dyon masses. However, the dyons also receive a positive mass-squared contribution from the Yukawa-like coupling to the meson field Eq.~(\ref{eq:monYuk}). The co-dependence of the Yukawa and gauge beta functions results in a flow to a fixed ratio between the two couplings. This ratio is such that the mass-squared due to loop AMSB is positive for both the meson and the dyons. Thus, the loop-level AMSB trilinear term in combination with the tree-level quartic potential gives a local minimum a distance $\mathcal{O}(\frac{m}{16\pi^2})$ from the origin. To understand the symmetry breaking pattern at this minimum, we must examine the form of the tree-level potential in terms of the $SU(N_F)$ matrix $M$ and vectors $q^\pm$,
\begin{eqnarray}
	V~&=&~\frac{1}{2} (q^+ \cdot q^{+*})(q^- \cdot q^{-*})\nonumber\\[3pt]
	&&+|M q^+|^2 + |M q^-|^2 + V_\text{AMSB}
\end{eqnarray}
The dot product term is due to the symmetric nature of the meson matrix ({\it i.e.}\/, $M^{ij}$ couples to $q_i^+ q_j^-$ and $q_j^+ q_i^-$). This term encourages the $q^\pm$ VEVs to point in different directions in flavor space. We find a minimum along the direction,
\begin{align}
	q^+ = \left(\begin{array}{c} 
		1 \\ 
		0 \\ \hline
		0 \\
		\vdots \\
		0
		\end{array} \right) \alpha,\quad
    q^- = \left(\begin{array}{c} 
		0 \\ 
		1 \\ \hline
		0 \\
		\vdots \\
		0
		\end{array} \right) \alpha,
\end{align}
\begin{align}
	M \propto \left( \begin{array}{cc|ccc}
		0 & 1 & 0 & \cdots & 0 \\
		1 & 0 & 0 & \cdots & 0 \\ \hline
		0 & 0 & 0 & \cdots & 0 \\
		\vdots & \vdots & \vdots & \ddots & \vdots \\
		0 & 0 & 0 & \cdots & 0
		\end{array} \right),
\end{align}
Breaking the global flavor symmetry to $SU(N_F-2)$. The vacuum energy of this minimum is $V=-\mathcal{O}(\frac{m}{16\pi^2})^4$.

In the vicinity of the singular point $U=U_1$, on the other hand, the light degrees of freedom are the monopoles $E^{\pm}\sim q^{\pm}_iQ^i$, whose magnetic charges are $\pm1$. These transform under the $SU(N_F)\times U(1)_R$ global symmetry of the UV theory as $E^{\pm}(\bf1)_1$. Since $\text{det}M\equiv U\neq0$, the global $SU(N_F)\times U(1)_R$ is spontaneously broken to $SO(N_F)\times U(1)_R$. In the neighborhood of $U=U_1$, the theory generates a dynamical superpotential
\begin{eqnarray}
W_{\text{mon}}\,&=&\,\tilde{f}\left(\frac{U-U_1}{\Lambda^{2N_F}}\right)\,E^+\,E^-\,.
\end{eqnarray}
Here $\tilde{f}(t)=t+\dotsb$ is holomorphic near $t=0$. For all practical purposes, only the leading order in $\tilde{f}$ matters for the stabilization of the minimum. Using canonically normalized fields we have
\begin{eqnarray}
W_{\text{mon}}\,&=&\,\Lambda\left(\frac{\tilde{U}}{\Lambda^{N_F}}-16\right)\,
\tilde{E}^+\,\tilde{E}^-\,,
\end{eqnarray}
where $\tilde{E}^\pm = E^\pm /\sqrt{\Lambda}$, $\tilde{U}=\text{det}\tilde{M}$, and $\tilde{M}=M/\Lambda$ is the canonically normalized meson. Exactly at $\tilde{U}=\tilde{U}_1\equiv16\Lambda^{N_F}$, 't Hooft anomaly matching is saturated by $E^{\pm},\,\tilde{M}^{ij}$, and the photinos $\mathcal{W}_\alpha\sim W_\alpha Q^{N_c-2}$, whose charges are given in Table~\ref{tab:Ncm2U1}.\\\quad\\
Explicitly, \\\quad\\
$U(1)_R\,{\rm gravity}^{2}$ and $U(1)^{3}_R:$
\begin{align}
	&(-1)N_FN_c+(1)\frac{N_c(N_c-1)}{2}= (1)+(-1)\frac{N_F(N_F+1)}{2},
\end{align}
$U(1)_R\,SO(N_F)^{2}$:
\begin{align}
(-1)(1)N_c&=(-1)(N_F+2).
\end{align}
\begin{table}[t]
	\centerline{
	\begin{tabular}{|c|c|c|c||c|c|} \hline
	& $SO(N_c)$ & $SU(N_F)$&$U(1)_R$&$U(1)_{\text{mag}}$&$SO(N_F)$ \\ \hline
	$Q^i$ & $\Yfund$ & $\Yfund$&$0$&$-$&$\Yfund$ \\ \hline 
	$\lambda$&$\Yasymm$&$\bf 1$&$1$&$-$&$\bf 1$\\ \hline\hline
	 $M_{ij}$&$\bf1$&$\Ysymm$&$0$&$-$&$\bf1+\Ysymm$\\ \hline
	 $E^{\pm}$&$-$&$\bf 1$&$1$&$\pm 1$&$\bf 1$\\ \hline
	 $\lambda_{\text{mag}}$&$-$&$\bf 1$&$1$&$0$&$\bf 1$\\ \hline
	\end{tabular}
	}
	\caption{DOF of the $SO(N_c)$ theory with $N_F=N_c-2$ near $M^{ij}\sim\delta^{ij}, U=U_1$. The unbroken global symmetry near $M^{ij}\sim\delta^{ij}, U=U_1$ is $SO(N_F)\times U(1)_R$, with $U(1)_R$ explicitly broken by AMSB.}\label{tab:Ncm2U1}
\end{table}

Contrary to the point $\tilde{U}=0$, here adding AMSB generates a tree-level contribution to the scalar potential from \eqref{eq:AMSBW}. This results in the global minimum at $\tilde{U}=\tilde{U}_1$. In particular, the scalar potential along $\tilde{M}^{ij}=\tilde{M}\delta^{ij}$ is given locally as
\begin{align}\label{eq:Ncm2pot}
\lefteqn{V_{\tilde{U}\sim \tilde{U}_1}\,=\,\Lambda^2\left|{\left(\frac{\tilde{M}}{\Lambda}\right)}^{N_F}-16\right|^2\left(|\tilde{E}^+|^2+|\tilde{E}^-|^2\right)}\nonumber\\
&+\frac{1}{kN_F}\left|N_F{\left(\frac{\tilde{M}}{\Lambda}\right)}^{N_F-1}\right|^2 |\tilde{E}^+\tilde{E}^-|^2\,+\,V_{\text{AMSB}}\,.
\end{align}
Note the $(k N_F)^{-1}$ factor in the second line, which comes from the K\"ahler term $k N_F \tilde{M}^\dagger \tilde{M}$ for $\tilde{M}$, where $k$ is an unknown $\mathcal{O}(1)$ normalization factor. The tree-level AMSB contribution is given by \eqref{eq:AMSBW}, {\it i.e.}\/,
\begin{eqnarray}\label{eq:Ncm2AMSB}
V_{\text{AMSB}}=m\Lambda\left[16+(N_F-1){\left(\frac{\tilde{M}}{\Lambda}\right)}^{N_F}\right]\,\tilde{E}^+\tilde{E}^-+\text{c.c.}\nonumber\\
\end{eqnarray}
This potential has a minimum at
\begin{eqnarray}\label{eq:globalNcm2}
&&\tilde{M}\,=\,16^{\frac{1}{N_F}}\Lambda~~~,~~~|\tilde{E}^+||\tilde{E}^-|=16^{\frac{2}{N_F}-1}\,km\Lambda\nonumber\\[5pt]
&&~~V_{\text{min}}\,=\,-16^{\frac{2}{N_F}}N_F\,km^2\Lambda^2\,.
\end{eqnarray}
Since $\tilde{M}^{ij}=\tilde{M}\delta^{ij}$ in this minimum, the global symmetry is broken to $SO(N_f)$, and there are no 't Hooft anomalies to match. Because it is generated by a tree-level contribution from AMSB, it is lower than the local minimum near the origin $U\approx 0$, and so it is the global minimum of the theory. A similar phenomenon of AMSB leading to a global minimum generated by a tree-level contribution and a local minimum generated at loop-level was seen in \cite{Csaki:2021aqv}. It is easy to see that there are no minima with $M^{ij}\neq0,\,M^{ij}\slashed{\propto}\delta^{ij}$.

Notably, the minimum \eqref{eq:globalNcm2} involves the condensation of monopoles $E^\pm$, which in turn leads to confinement \cite{tHooft:1981bkw,Cardy:1981qy,Cardy1982}, in the sense that non-trivial Wilson lines get an area law. Famously, monopole condensation has also been seen in the breaking of $\mathcal{N}=2$ Seiberg-Witten theory to $\mathcal{N}=1$ via a tree level superpotential for the matter field \cite{Seiberg:1994rs}. In \cite{Evans:1996hi,Konishi:1996iz}, it was shown in a non-supersymmetric theory by introducing soft SUSY breaking on top of the breaking to $\mathcal{N}=1$. Here, on the other hand, monopole condensation and SUSY breaking emerge together as a result of AMSB. Furthermore, since the global $SU(N_F)$ symmetry is spontaneously broken to $SO(N_F)$, this is a demonstration of \textit{confinement with chiral symmetry breaking} in a non-supersymmetric setting.

We can also connect the chiral symmetry breaking observed here to the familiar one due to fermion bilinears. To see this, note that the UV theory of quarks has no superpotential and their F-components vanish. Therefore the only contribution to the $F$-component of the \textit{meson} superfield comes from fermion bilinears:
\begin{align}
\langle \psi_{i}^* \psi_{j}^* \rangle = F^*_{M_{ij}} = 16 \Lambda^2 M^{-1}_{ij} E^{+} E^{-} \propto \delta_{ij} k m \Lambda^{2} \neq 0.
\end{align}
In other words, our analysis demonstrates the condensation of fermion bilinears in a non-supersymmetric theory, in addition to the monopole condensate.

\subsection{$N_F<N_c-4$}
At a generic point in the moduli space, the gauge group $SO(N_c)$ is higgsed down to $SO(N_c-N_F)$ pure SYM, whose gaugino condensation induces an Affleck--Dine--Seiberg (ADS) superpotential given by
\cite{Affleck:1983vc}
\begin{align}\label{eq:ADS}
W_{\text{ADS}}&=\frac{N_c-N_F-2}{2}\omega^k\left(\frac{16\Lambda^{3N_c-N_{F}-6}}{\text{det} M}\right)^{\frac{1}{ N_{c}-N_{F}-2}}\,,
\end{align}
where $\Lambda$ is the strong scale of the theory and $\omega = e^{2\pi i/(N_c-N_F-2)}$ with $k=0, 1, \dots, N_c-N_F-3$. 

The K\"ahler potential of $M$ is singular at the origin and writing $M^{ij}=\varphi^2\delta^{ij}$, we identify $\varphi$ as the canonical DOF. Turning on AMSB stabilizes the runaway behavior of the superpotential via the tree-level scalar potential
\begin{equation}\label{eq:NFlNcm4AMSB}
V_{\text{AMSB}}=-m\Lambda^3\,\frac{3N_c-N_F-6}{2}\left(\frac{16\Lambda^{2N_{F}}}{\varphi^{2N_F}}\right)^{\frac{1}{ N_{c}-N_{F}-2}}+\text{c.c.}\,,
\end{equation}
which together with the scalar potential derived from the superpotential \eqref{eq:ADS} gives a minimum
\begin{eqnarray}\label{eq:globalNFlNcm4}
\varphi&=&~~2^{\frac{2}{N_c-2}}\,{\left(f_{N_F}\frac{\Lambda}{m}\right)}^{\frac{N_c-N_F-2}{2(N_c-2)}}~\Lambda\nonumber\\
V_{\text{min}}&=&-2^{\frac{4}{N_c-2}}\frac{N_c-2}{f^2_{N_F}}{\left(f_{N_F}\frac{\Lambda}{m}\right)}^{\frac{N_c-N_F-2}{N_c-2}}~m^2\Lambda^2\,.\nonumber\\
\end{eqnarray}
with $f_{N_F}=\frac{N_c+N_F-2}{3N_c-N_F-6}$.  
We see that the minimum is at $\varphi\gg\Lambda$, which justifies a weakly coupled analysis in an asymptotically free theory. Since $\tilde{M}^{ij}\propto\delta^{ij}$, in this minimum the global symmetry is broken to $SO(N_F)$. There are no minima with $\tilde{M}^{ij}\neq0,\,\tilde{M}^{ij}\slashed{\propto}\delta^{ij}$. Since the $U(1)_R$ symmetry was explicitly broken by AMSB, there are no 't Hooft anomaly matching conditions to check in this scenario.

The non-trace components of $\tilde{M}^{ij}$ are split into massless Nambu--Goldstone bosons (NGBs), massive fermions, and massive scalar partners of the NGBs, where masses are $\mathcal{O}(m)$. The NGBs form the chiral Lagrangian on the $SU(N_F)/SO(N_F)$ coset space. Once the massive fermions are integrated out, the one-loop diagram \cite{DHoker:1984izu} produces the Wess--Zumino--Witten (WZW) term \cite{Wess:1971yu,Witten:1983tw} because $\pi_5 (SU(N_F)/SO(N_F)) = {\mathbb Z}$ ($N_F\geq 3)$. For $N_F=2$, there is no WZW term. To summarize, we establish that the $1< N_F<N_c-4$ case with AMSB has a global minimum in which the chiral symmetry is broken to $SO(N_F)$, similar to the $N_F=N_c-2$ case.

The case $N_F=1$ is an exception because the meson has only one component. There is no exact flavor symmetry, no massless NGB, and the theory is gapped.

\subsection{$N_F=N_c-4$}
In this case the gauge symmetry is higgsed on the moduli space to $SO(4)\simeq SU(2)_L\times SU(2)_R$. The theory has two distinct branches corresponding to the gaugino condensates in $SU(2)_{L,R}$ having the same, or opposite signs. On the first branch with aligned condensates, the superpotential is of the same form as \eqref{eq:ADS}, while on the second branch, it vanishes. The second branch contains the point $M=0$, at which there is confinement without chiral symmetry breaking. In \cite{Csaki:1997aw}, it was shown that on this branch there is also a VEV for the exotic baryon $S=W_\alpha W^\alpha Q^{N_F}$, which breaks the discrete global symmetry $\mathbb{Z}_{2F}$ down to $\mathbb{Z}_F$.

With AMSB, the theory on the first branch develops a minimum identical to \eqref{eq:globalNFlNcm4}, breaking the global symmetry down to $SO(N_F)$. This is the global minimum of the theory.
As for the second branch, the identically zero superpotential means we need to consider a more general version of Eq.~\eqref{eq:AMSBW} that accounts for a non-canonical K\"ahler potential. We find for a general $W$ and $K$
\begin{align}\label{eq:extendedAMSB}
	V_{\rm tree} = & \partial_i W g^{i j^*} \partial_j^* W^*  
	+ m^* m \left( \partial_i K g^{i j^*} \partial_j^* K - K \right) \nonumber \\
	& + m \left(\partial_i W g^{i j^*} \partial_j^* K - 3 W \right) + c.c. ,
\end{align}
where $g^{ij}$ is the inverse of the K\"{a}hler metric $g_{ij} = \partial_i \bar{\partial}_j K$. With a vanishing superpotential, interactions originate only from the K\"{a}hler potential. Higher order terms in the K\"{a}hler potential will give rise to irrelevant interactions, and as the theory is IR free, we expect the effects of AMSB to be highly suppressed by the dynamical scale. To estimate these effects, consider the leading corrections to the canonical K\"{a}hler potential for $M$,
\begin{eqnarray}
K~&=&~\mathrm{Tr}{M^\dagger M} + \frac{a}{\Lambda^2} \left(\mathrm{Tr}{M^\dagger M}\right)^2\nonumber\\
&&+ \frac{b}{\Lambda^2} \mathrm{Tr}{M^\dagger M M^\dagger M}\, ,
\end{eqnarray}
where $a$, $b$ are order one numbers. Note that cubic terms are forbidden because $M$ is in the symmetric of $SU(N_F)$.

In this case, only the second term in Eq.~\eqref{eq:extendedAMSB} will contribute, and at leading order gives,
\begin{align}
	V &\sim \pm\frac{m^2}{\Lambda^2} |M|^4 
\end{align}

The potential in this theory arises exclusively from AMSB. Clearly the power series expansion makes sense only up to $M \sim \Lambda$, and the maximum contribution of the higher dimension terms to the potential is $O(m^2 \Lambda^2)$. Note that the minimum we obtained in Eq.~\eqref{eq:globalNFlNcm4} (and is also relevant for the branch with nonzero superpotential) is parametrically enhanced by $(\Lambda/m)^{2/(N_c-2)}$. Therefore the branch with $W=0$ does not yield the global minimum, which instead arises from the branch with the ADS-type superpotential. 

\subsection{$N_F=N_c-3$}
Here the gauge symmetry on the moduli space is higgsed down to $SO(3)$. As in \cite{Intriligator:1995id}, it is useful to first turn on the VEVs of $N_F-1$ of the fundamentals, in which case the theory is higgsed to $SU(2)_L\times SU(2)_R$. Then, the VEV of the last fundamental higgses $SU(2)_L\times SU(2)_R$ to the diagonal $SO(3)$. The superpotential for this theory is dynamically generated by a combination of gaugino condensation in the unhiggsed $SO(3)$ and instantons in the broken $SU(2)_{L}\times SU(2)_{R}/SO(3)$. The theory again has two branches: on the first, the gaugino contribution is aligned with the instanton contributions, and the superpotential is of the form \eqref{eq:ADS}. On the second branch, the contributions cancel out, and the ADS-type superpotential vanishes. However, it was shown via a mass deformation to the $N_F=N_c-4$ case that the second branch has a dynamically generated superpotential:
\begin{eqnarray}
W_{\text{dyn}}\,&=&\,\frac{1}{2\mu}\,f(t)\,M^{ij}q_{i}q_{j}\,,
\end{eqnarray}
where $q_i=\left(W_\alpha W^\alpha Q^{N_F-1}\right)_i/\Lambda^{N_F+1}$ is the exotic baryon of the theory. Here $\mu$ is an effective mass scale, $t=\frac{1}{\Lambda^{2N_F+4}}\text{det}M\,M^{ij}q_{i}q_{j}$, and $f(t)$ is a holomorphic function in the neighborhood of $t=0$, normalized so that $f(0)=1$. We can canonically normalize the superpotential yielding
\begin{eqnarray}
W_{\text{dyn}}\,&=&\, \frac{1}{2} f(t)\,\tilde{M}^{ij}q_{i}q_{j}\,,
\end{eqnarray}
where $\tilde{M}$ is the canonically normalized meson.

As usual, adding AMSB to the theory generates for the first branch the minimum \eqref{eq:globalNFlNcm4} and breaks the global symmetry down to $SO(N_F)$. This is again the global minimum of the theory.
On the second branch, the tree-level AMSB contribution vanishes at $\mathcal{O}(t^0)$, while the loop-level contribution \eqref{eq:AMSBloop} generates a minimum in the neighborhood of $\tilde{M}^{ij}=0$, with 
\begin{align}\label{eq:notglobal}
V\,&\approx\,-\left(\frac{m}{16\pi^2}\right)^4\,,
\end{align}
Again the $\mathcal{O}(t)$ corrections give a sub-leading contribution. Around the origin loop-level AMSB is again the dominant perturbation, which will  clearly not be the global minimum of the theory, since its (negative) height is loop suppressed.

\subsection{$ N_c-1\leq N_F\leq \frac{3}{2}(N_c-2)$}
In this case the correct IR description of the theory is in terms of its IR free Seiberg dual  $SO(N_F-N_c+4)$ with $N_F$ fundamentals $q_i$ and $\frac{1}{2}N_F(N_F+1)$ singlets $M^{ij}$ in the $q_i(\overline{\Yfund})_{\frac{N_c-2}{N_F}}$ and $M^{ij}(\Ysymm)_{\frac{2(N_F-N_c+2)}{N_F}}$ representations of the global $SU(N_F)\times U(1)_R$, respectively.
First we will focus on the case when $N_F\geq N_c+1$, leaving the $N_F=N_c-1$ and $N_F=N_c$ special cases to the end of the section. For $N_F\geq N_c+1$, the dual theory has a superpotential 
\begin{eqnarray}\label{eq:supmag}
W_{\text{dual}}\,&=&\,\frac{1}{2\mu}\,M^{ij}q_iq_j\,.
\end{eqnarray}
The scales of the original and dual theories are related by 
\begin{equation}\label{eq:relmag}
2^8\Lambda^{3(N_c-2)-N_F}\tilde{\Lambda}^{2N_F-3(N_c-2)}~=~(-1)^{N_f-N_c}\mu^{N_F}\,. 
\end{equation}
For later convenience, we switch to canonically normalized fields $\tilde{M}$,
\begin{eqnarray}\label{eq:supmag}
W_{\text{dual}}\,&=&\,\frac{1}{2}\,\tilde{M}^{ij}q_iq_j\,,
\end{eqnarray}
with $\tilde{M}=M/\mu$.
When we turn on AMSB, the situation is similar to the one encountered in \cite{Csaki:2021aqv} and to the $N_F=N_c-2$ case in the present work. Near $M=0$ there is a local minimum generated by the loop level AMSB contribution. The tree-level contribution vanishes as usual because the superpotential \eqref{eq:supmag} is marginal. Again we expect only a local minimum with $V=-\mathcal{O}(\frac{m}{16\pi^2})^4$ along the direction,
\begin{align}
	q \propto \left(\begin{array}{ccc} 
		1 & \cdots & 0 \\ 
		\vdots &\ddots & \vdots \\
		0 & \cdots & 1 \\ \hline
		0 & \cdots & 0 \\
		\vdots & \ddots & \vdots \\
		0 & \cdots & 0
		\end{array} \right), &\quad
	M \propto \left( \begin{array}{ccc|ccc}
		1 & \cdots & 0 & 0 & \cdots & 0 \\
		\vdots & \ddots & \vdots & \vdots & \ddots & \vdots \\
		0 & \cdots & 1 & 0 & \cdots & 0 \\ \hline
		0 & \cdots & 0 & 0 & \cdots & 0 \\
		\vdots & \ddots & \vdots & \vdots & \ddots & \vdots \\
		0 & \cdots & 0 & 0 & \cdots & 0 
		\end{array} \right)\, ,
\end{align}
where $q$ is a $N_F\times (N_F-N_c+4)$ matrix.

At nonzero $M$ with $\text{rank}(M)=N_F$ on the moduli space for $N_F\geq N_c+1$, the dual quarks $q_i$ are integrated out, and we are left with pure $SO(N_F-N_c+4)$ SYM, with a scale
\begin{eqnarray}\label{eq:Wdynmag}
\tilde{\Lambda}_{L}^{3N_F-3(N_c-2)}\,&=&\, \text{det}(\tilde{M})~\tilde{\Lambda}^{2N_F-3(N_c-2)}
\end{eqnarray}
Gaugino condensation in the dual theory now generates a dynamical superpotential
\begin{eqnarray}\label{eq:Wdynmag}
W_{\lambda\lambda}\,=\,2^{-\frac{N_F-(N_c-2)-4}{N_F-(N_c-2)}}\epsilon_{N_F-(N_c-2)}\,\,\tilde{\Lambda}^3_{L}
\end{eqnarray}
With AMSB this superpotential leads to the tree-level SUSY breaking scalar potential \begin{eqnarray}\label{eq:NFlNcm4AMSB}
V_{\text{AMSB}}=&&-2m\tilde{\Lambda}^3\,\frac{\frac{3}{2}(N_c-2)-N_F}{N_F-(N_c-2)}\times\\
&&\left(\frac{16^{\frac{1}{N_F}}\text{det} (\tilde{M})}{\tilde{\Lambda}^{N_{F}}}\right)^{\frac{1}{N_{F}-(N_{c}-2)}}+c.c.\nonumber
\end{eqnarray}
For $N_F>N_c$, this potential, together with the usual scalar potential from the superpotential, is minimized at
\begin{eqnarray}\label{eq:globallNc}
\tilde{M}^{ij}&\sim&4^{\frac{N_F-(N_c-2)-2}{2(N_c-2)-N_F}}\left(f\frac{m}{\Lambda}\right)^{\frac{N_F-(N_c-2)}{2(N_c-2)-N_F}}\,\Lambda\,\delta^{ij}\nonumber\\[5pt]
V_{\text{min}}&\sim&-4^{\frac{N_F-4}{2((N_c-2)-N_F}}\frac{2(N_c-2)-N_F}{[N_F-(N_c-2)]^2}\times\nonumber\\
&&\left(f\frac{m}{\Lambda}\right)^{\frac{2(N_c-2)}{2(N_c-2)-N_F}}\,\Lambda^4\,,
\end{eqnarray}
where $f=\frac{1}{N_c-2}\left[N_F-(N_c-2)\right]\left[\frac{3}{2}(N_c-2)-N_F\right]$. Noting that $N_c-2<N_F\leq\frac{3}{2}(N_c-2)$ from \eqref{eq:globallNc}, this minimum satisfies $\tilde{\Lambda}_L\ll \tilde{M}\ll \tilde{\Lambda}$, below the Landau pole of the UV magnetic theory and above the scale of gaugino condensation in the IR pure SYM. This justifies our weakly coupled analysis. Again the global symmetry at the minimum is broken down to $SO(N_F)$.

We now comment on the $N_F=N_c-1$ and $N_F=N_c$ cases. For $N_F=N_c-1$, the dual gauge group is $SO(3)$. The superpotential of the dual now includes a contribution from instantons:
\begin{eqnarray}\label{eq:supmagso3}
W_{\text{dual},N_c-1}\,&=&\,\frac{1}{2\mu}\,M^{ij}q_iq_j\,-\,\frac{\text{det}M}{64\tilde{\Lambda}^{2N_c-5}}\,.
\end{eqnarray}
The extra contribution can be seen by deforming the dual for $N_F=N_c$ by a mass term for the last flavor, which leads to instantons in the broken $SU(2)_L\times SU(2)_R/SO(3)$ in the magnetic theory \cite{Csaki:1998vv}. Here the scale matching is given by
\begin{equation}
2^{14}\,\left(\Lambda^{2N_c-5}\,\tilde{\Lambda}^{4-N_c}\right)^2~=~\left(\mu^{N_c-1}\right)^2\,. 
\end{equation}
Interestingly, this relation looks like the square of the usual relation \eqref{eq:relmag} - for more details, see the original \cite{Intriligator:1995id}. When $M$ has full rank on the moduli space, the $q_i$ become massive and the gaugino condensation in the IR $SO(3)$ SYM generates a superpotential
\begin{eqnarray}\label{eq:supmagso3la}
W_{\lambda\lambda,\text{dual},N_c-1}\,&=&\,\epsilon_2\,\frac{\text{det}M}{64\Lambda^{2N_c-5}}\,.
\end{eqnarray}
Note that this superpotential has the same magnitude as the instanton contribution in \eqref{eq:supmagso3}, and so the theory again has two branches: one in which the two contributions add, leading to an AMSB minimum of the form \eqref{eq:globallNc}, and another with vanishing superpotential. On the second branch there is no superpotential, and we can repeat the arguments concerning the second branch in the case of $N_F=N_c-4$. Again, any minimum produced by AMSB and a non-canonical K\"ahler is parametrically suppressed in comparison with the first branch.

For $N_F=N_c$, the magnetic gauge group is $SU(2)_L\times SU(2)_R$. There is no instanton contribution to the tree-level superpotential, but when $M$ is given full rank, there are again two branches: one with aligned gaugino condensates in the IR pure $SU(2)_L\times SU(2)_R$ SYM, and one with opposite sign condensates and vanishing superpotential. As usual, with AMSB the first branch leads to a global minimum of the form \eqref{eq:globallNc}, while any minimum in the second branch is again subdominant.

Finally, we note that for $N_c< 6$ we have $\frac{3}{2}(N_c-2)< N_c$, and so the theories with $N_F=N_c=4,5$ have to be considered separately. Indeed, for $N_F=N_c=4,5$ the magnetic theory is no longer IR free, but rather has an IR fixed point. 

\subsection{Monopole Condensation for $N_F<N_c-2$\\ via Mass Deformations}

When discussing the theory with $N_F=N_c-2$, the non-supersymmetric vacuum of the theory explicitly involved monopole condensation. In the next section, this will enable us to determine the behavior of the loop operators in the theory, and in particular establish confinement of non-trivial Wilson loops for $Spin(N_c)$. In this section we wish to make contact between the $N_F=N_c-2$ case and the cases with fewer flavors, by treating the latter as the $N_F=N_c-2$ deformed by a supersymmetric mass $\mu$, with $\mu\gg\Lambda$. 

\begin{figure}[ht]
\begin{center}
\includegraphics[width=0.9\linewidth]{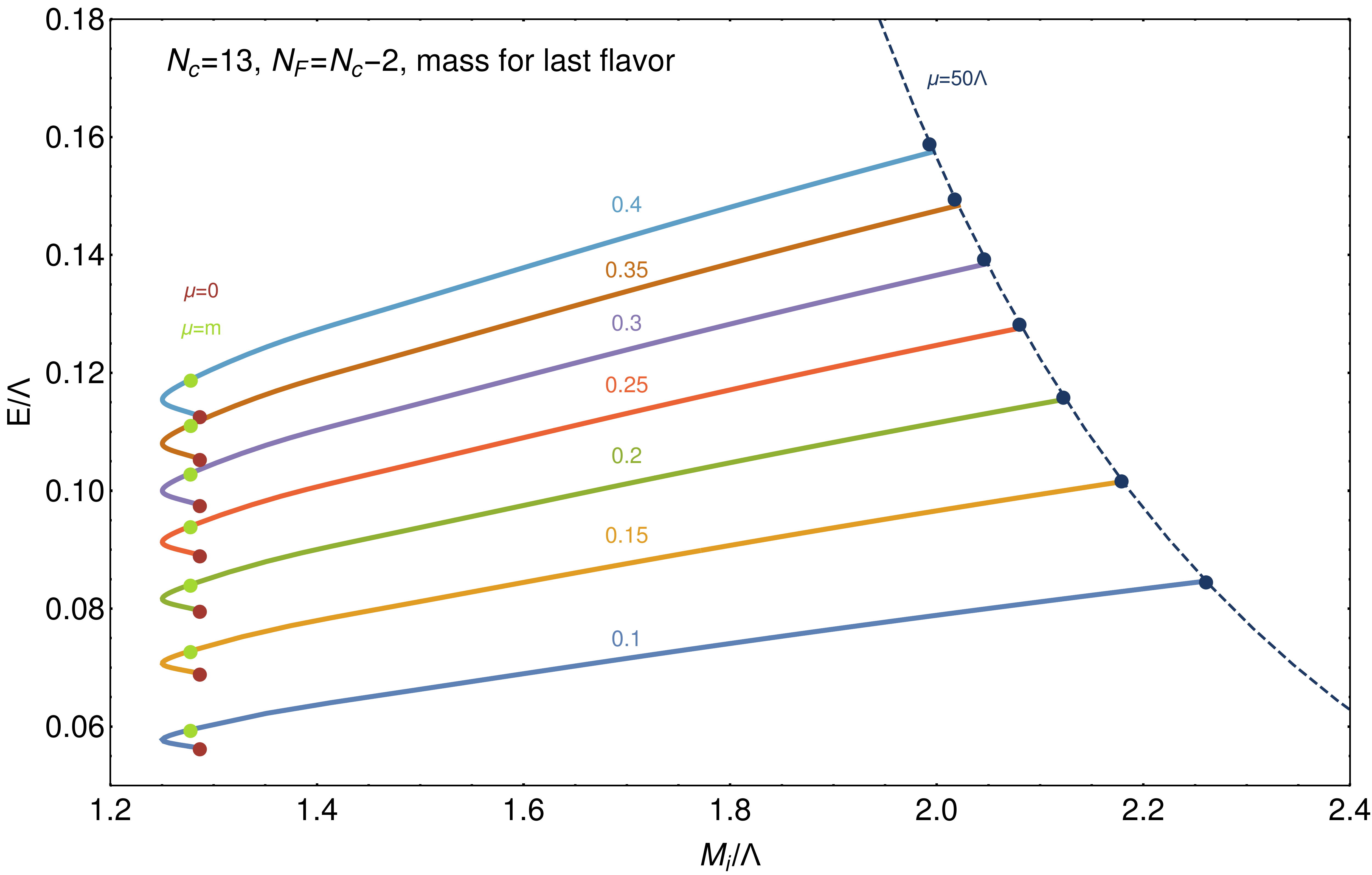}
\caption{The supersymmetry breaking minimum for the theory with AMSB and $N_F=N_c-2$, with the extra mass term $\frac{1}{2}\mu M^{N_FN_F}$. $E$ is the VEV of the monopoles $\tilde{E}^\pm$, while $M_{i}$ is the VEV of the first $N_c-3$ flavors. The labeling on the different curves indicates different values of $m/\Lambda$. For $\mu=0$, the curves are at the $N_c-2$ minimum \eqref{eq:globalNcm2} .  As $\mu$ grows, the VEV of $M_i$ initially decreases, but then starts increasing as $\mu$ passes $m$. For large $\mu/\Lambda\rightarrow \infty$, the vacuum of the theory goes over to  Eq.~\eqref{eq:globalNFlNcm4} with $N_F=N_c-3$, while the monopole condensate persists. The relation \eqref{eq:EElower} is shown in the dashed line for $\mu=50\Lambda$. We chose $N_c=13$ for this plot.}\label{fig:mudef1}
\end{center}
\end{figure}

We begin by considering the $N_F=N_c-2$ theory in the supersymmetric limit, with just one mass term for the last flavor,
\begin{eqnarray}\label{eq:Ncm2mass}
W&=&\Lambda\left(\frac{\det \tilde{M}}{\Lambda^{N_F}}-16\right)\,\tilde{E}^+\,\tilde{E}^- + \frac{1}{2}\mu \Lambda \tilde{M}_{N_F N_F}
\end{eqnarray}
The equation of motion for $\tilde{M}_{N_F N_F}$ gives
\begin{eqnarray}\label{eq:EElower}
\tilde{E}^+ \tilde{E}^- \,&=&\, -\frac{1}{2} \frac{\mu \Lambda^{N_F}}{\det \tilde{M}'}\, ,
\end{eqnarray}
where $\tilde{M}'$ is the matrix of the remaining mesons. As already demonstrated, tree level AMSB corrections to the ADS superpotential stabilize the runaway, and the finite VEV of $M'$ ensures that a non-vanishing monopole condensate persists. In Fig.~\ref{fig:mudef1} we show this explicitly by studying the minimum of the mass-deformed theory ~\eqref{eq:Ncm2mass} in the presence of AMSB with $m<\mu$. Since we are ultimately interested in the infinite $\mu$ limit, this does not interfere with our extrapolation to the non-SUSY limit with large $m$.
As can be seen in the plot, the VEV of the first $N_c-3$ flavors interpolates between the minimum \eqref{eq:globalNcm2} for $\mu=0$, and the ADS+AMSB minimum \eqref{eq:globalNFlNcm4} with $N_f=N_c-3$ and $\Lambda\rightarrow\Lambda_{N_F=N_c-3}$ in the large $\mu$ limit. We can see that the monopole condensate persists in the large $\mu$ limit. 

To correctly reproduce the ADS+AMSB minimum, we had to interpolate the K\"ahler potential between the neighborhood of $\text{det} \tilde{M} \sim \tilde{U}_1$, where it is canonical in $\tilde{M}$, to large $\text{det} \tilde{M}$, where the K\"ahler potential is canonical in $\varphi\sim\sqrt{\tilde{M}\Lambda}$. More specifically, we used the following interpolating K\"ahler potential in the numerical study:
\begin{eqnarray}\label{eq:Kinterp}
K_{\text{interp.}}~=~\Lambda^2\,\sqrt{1+\frac{\tilde{M}\tilde{M}^\dagger}{\Lambda^2}}\,.
\end{eqnarray}
For $\mu<m$, the UV theory has a runaway at $E^+E^-=0$ and $M_i\rightarrow\infty$. This is a consequence of the mass term in \eqref{eq:Ncm2mass} in the presence of AMSB. In this regime we follow the local minimum which goes over to the global minimum for $\mu>m$. Importantly, the condensation of monopoles in the large $\mu$ limit is independent of this subtlety.

\begin{figure}[ht]
\begin{center}
\includegraphics[width=0.9\linewidth]{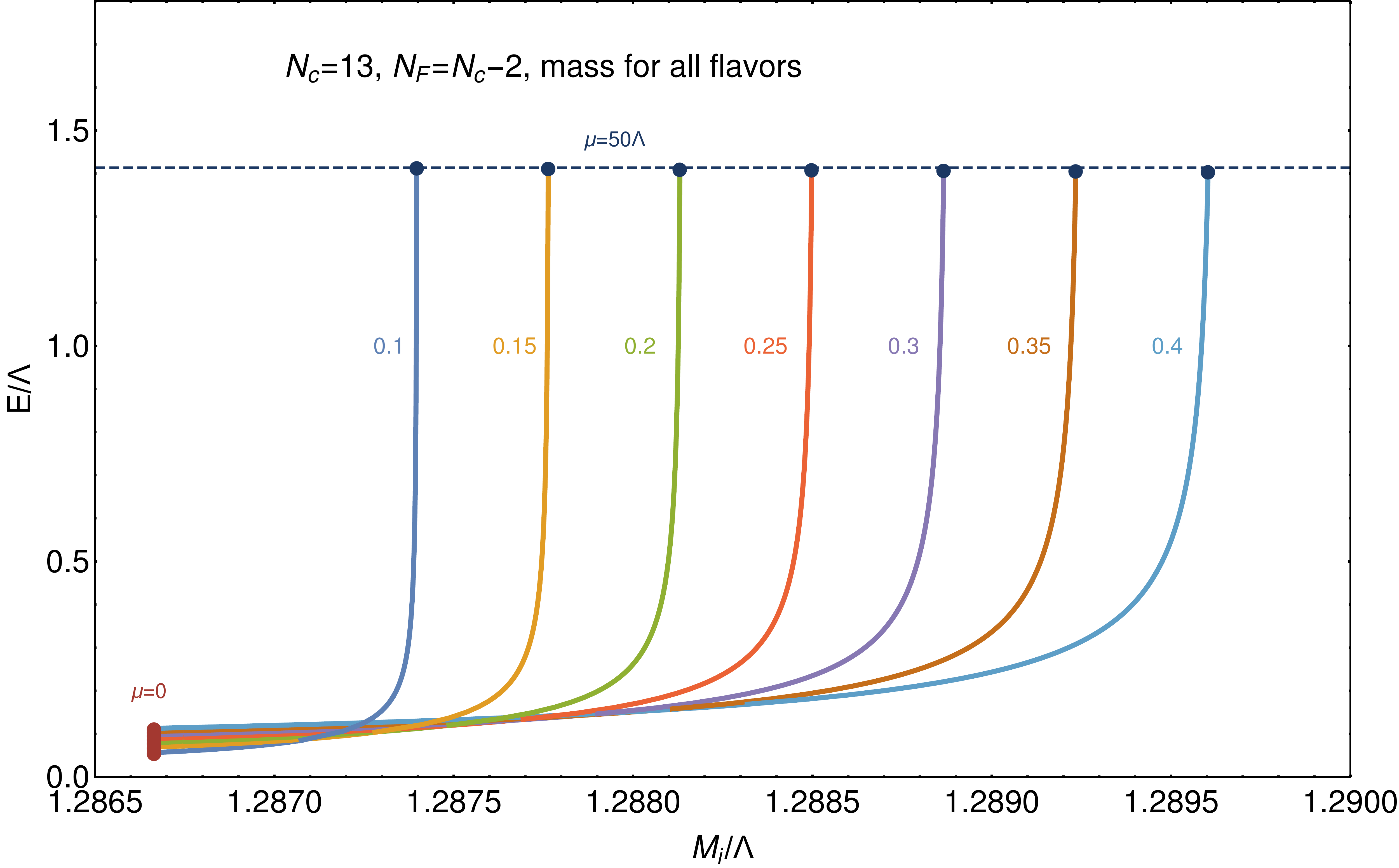}
\caption{Location of the minimum in the theory with AMSB and $N_F=N_c-2$, deformed by a universal mass term $\frac{1}{2}\mu M^{ii}$ for all flavors. $E$ is the VEV of the condensed monopoles, while $M_{i}$ is the common VEV for all of the flavors. The different curves are labeled by the value of $m/\Lambda$. The curves start at the $N_c-2$ minimum \eqref{eq:globalNcm2} for $\mu=0$.  As $\mu/\Lambda\rightarrow \infty$, the theory goes over to pure SYM, while the VEV of the monopoles $E$ is given by \eqref{eq:EElowerSYM}, represented by the dashed line. We have again chosen   set $N_c=11$ for this plot.}\label{fig:mudef2}
\end{center}
\end{figure}

Similarly, we can give a mass term to any number of flavors in the $N_F=N_c-2$ theory and show that monopole condensation persists. In Fig.~\ref{fig:mudef2} we present the case where all of the flavors get the same mass term, resulting in a pure SYM theory with monopole condensation. In the small $\mu$ limit, the minimum is given by the $N_F=N_c-2$ vacuum \eqref{eq:globalNcm2}. The $\mu\gg m$ case can be fully understood in the supersymmetric limit -- the monopoles get a VEV
\begin{eqnarray}\label{eq:EElowerSYM}
\tilde{E}^+ \tilde{E}^-\,&=&\, -\frac{\mu\Lambda}{2} {\left(\frac{ \Lambda}{\tilde{M}}\right)}^{N_F-1}\,,
\end{eqnarray}
where $\tilde{M}$ is the common VEV of all of the flavors. This generates an ADS superpotential for the $\tilde{M}$, which is balanced by the $\mu$ term and leads to an overall minimum at
\begin{eqnarray}\label{eq:EElowerSYM}
\tilde{M} \,&=&\, 16^{\frac{1}{N_F}}\Lambda\,,\nonumber\\[3pt]
\tilde{E}^+ \tilde{E}^-\,&=&\, -2^{\frac{4}{N_F}-5}\,\mu\Lambda\,.
\end{eqnarray}
Notably, the VEV of $\tilde{M}$ in this case is equal to the pure $N_F=N_c-2$ case. This vacuum is the one depicted by the dashed line in Fig.~\ref{fig:mudef2}.

Though we don't show this explicitly, the same conclusion persists for any number of flavors that are integrated out from the $N_F=N_c-2$ theory, and so we explicitly see monopole condensation for the entire range $0\leq N_F\leq N_c-2$. 

\subsection{Loop Operators and Confinement}

The observed monopole condensation for $N_F \leq N_c-2$ implies the confinement of electric and dyonic loop operators and a perimeter law for magnetic 't Hooft loops. This is in agreement with arguments made in \cite{Aharony:2013}. For $N_F<N_c-4$, the VEV of the meson field Higgses the gauge group to pure YM with more than four colors, in which Wilson and dyonic loops confine while 't Hooft loops do not.

With $N_c-4$ flavors, the situation is a bit more subtle because the unbroken $SO(4)\simeq SU(2)_L\times SU(2)_R$ gauge theory forms a gaugino condensate for each $SU(2)$ factor. The branch where the gaugino condensates are aligned is connected by a mass deformation to the $N_c-2$ case, and so by the argument of the previous section it involves monopole condensation. Consequently, magnetic loops acquire a perimeter law while dyonic loops acquire an area law. The other branch with anti-aligned condensates is related to the first one by a shift $\theta_1\rightarrow\theta_1+2\pi$ \cite{Aharony:2013}, and so by the Witten effect, on this branch it is the dyons that condense. On this branch, dyonic loops acquires a perimeter law while magnetic loops acquires an area law. In both cases the electric Wilson loop is confined.

In the case of $N_c-3$ flavors, the unbroken gauge group is again special because $SO(3)_+$ is in fact related to $SO(3)_-$ by a shift of the vacuum angle $\theta \rightarrow \theta +2\pi$ \cite{Aharony:2013}. Such a shift permutes the magnetic and dyonic loops, however the shift also exchanges the two orientations of the gaugino condensate, thus exchanging the ADS and exotic baryon branches. Again the global minimum exhibits the same loop operator behavior as of all other values of $N_F\leq N_c-2$. We are free to interpret this case as either monopoles condensing in $SO(3)_-$ or dyons condensing in $SO(3)_+$. 

For $N_c-1\leq N_F\leq \frac{3}{2}(N_c-2)$, the global minimum of the dual theory finds the meson with non-vanishing VEV. Integrating out the dual quarks leaves behind pure YM, for which we already demonstrated monopole condensation in the previous section. Thus, the electric and dyonic loops of the \textit{dual theory} confine. Matching the behavior of the single non-trivial loop in each of the dual theories, the correspondence of gauge groups is \cite{Aharony:2013,Gaiotto:2014kfa}, 
\begin{eqnarray}\label{eq:freemagdual}
Spin(N_c) \longleftrightarrow SO(N_F-N_c+4)_- \nonumber\\
SO(N_c)_+ \longleftrightarrow SO(N_F-N_c+4)_+ \\
SO(N_c)_- \longleftrightarrow Spin(N_F-N_c+4) \nonumber
\end{eqnarray}
Put concisely, the duality exchanges the electric and dyonic loops. The duality (\ref{eq:freemagdual}) therefore implies the same loop operator behavior as in the $N_F\leq N_c-2$ case.

As first noted in \cite{Intriligator:1995id} and elaborated in \cite{Aharony:2013}, the cases $N_F=N_c-1,\,N_c$ are special in the sense that there are extra dual descriptions for the same theory. Let us first focus on the $N_F=N_c-1$ case, and take the original theory to be $Spin(N_c)$. In that case the dual is $SO(3)_-$ with $N_f=N_c-1$ flavors and a superpotential \eqref{eq:supmagso3}. But we know that this theory is equivalent to $SO(3)_+$ with $\theta$ shifted by $2\pi$, which results in an exchange of the ADS and the exotic baryon branch. Dualizing back, we find another IR description of the theory in terms of $SO(N_c)_+$ and a superpotential $W_{\text{2nd dual}}\,=\,-\,\frac{\text{det}M}{32\Lambda^{2N_c-5}}$. By the arguments above, at the AMSB minimum the dyonic loop of the $SO(3)_-$ description and the Wilson loop of the $Spin(N_c)$ description confine, while the 't Hooft loop of the $SO(N_c)_+$ description has a perimeter law. A similar logic applies if we choose the original theory to be $SO(N_c)_-$, in which case the first dual is $Spin(3)$ and the second dual is again $SO(N_c)_-$. Here both the Wilson loop of $Spin(3)$ and the dyonic loop of $SO(N_c)_-$ confine.
For $N_F=N_c$ there are again two dual descriptions of the original theory, albeit with a different superpotential for the second dual. Since the loop behavior is identical to the $N_F=N_c-1$ case, we do not repeat the analysis here.

In summary, theories with $N_F\leq \frac{3}{2}(N_c-2)$ and AMSB all experience monopole condensation and the same behavior for their loop operators. In particular, the non-trivial Wilson loop has an area law, signaling electric confinement.

\section{Non-Confining Phases}

For the supersymmetric theories with $ \frac{3}{2}(N_c-2)< N_F\leq 3(N_c-2)$, there is an IR fixed point, while for $3(N_c-2)<N_F$ the theories are IR free. Below we explore the behavior of the theory in these ranges when we add AMSB.

\subsection{$ \frac{3}{2}(N_c-2)< N_F\leq 3(N_c-2)$}
In this regime the theory has an IR fixed point. The IR dynamics is described by either the electric or the magnetic theory. At the fixed point, the electric and magnetic degrees of freedom pick up an anomalous dimension $\gamma_i=3R_i-2$, where $R_i$ is the $R$-charge. Since there is no additional anomaly-free global $U(1)$ symmetry, the $R$-symmetry in this case is uniquely defined. The anomalous dimensions are then:
\begin{eqnarray}\label{eq:anomdim}
\gamma_{Q}~&=&~\frac{N_F-3N_c+6}{N_F}\,,\nonumber\\
\gamma_{q}~&=&~\frac{3N_c-2N_F-12}{N_F}\,,\nonumber\\
\gamma_{M}~&=&~\frac{4N_F-6N_c+12}{N_F}\,.
\end{eqnarray}
It's easy to see that these anomalous dimensions are consistent with the vanishing of the NSVZ beta function. With AMSB, the theory becomes supersymmetric again at the IR fixed point, as in \cite{Murayama:2021xfj,Csaki:2021aqv}. This is reminiscent of the IR restoration of supersymmetry presented in \cite{Lanzagorta:1995ai,Sundrum:2009gv}. At intermediate scales, the gaugino mass is power-suppressed, and approaches zero quickly in the IR \cite{Murayama:2021xfj,Csaki:2021aqv}.
\subsection{$3(N_c-2)< N_F$}
Here the theory is in the free-electric phase, and so there is no superpotential. With AMSB, the squarks get a negative soft mass from \eqref{eq:AMSBloop} and become tachyonic. The theory then has no ground sate. The theory is then not continuously connected to non-SUSY $SO(N_c)$ with $3(N_c-2)< N_F$ fundamentals.

\section{Conclusions}

We have examined the low-energy phase structure of $SO(N_c)$ gauge theories with $N_F$ Weyl fermions in the vector representation, obtained by perturbing the SUSY version of this theory via AMSB. We found that the intricate phase structure of the SUSY theory does not survive the non-SUSY perturbation. Instead the phase structure is very simple: for small number of flavors $N_F\leq\frac{3}{2}(N_c-2)$ the theory is confining with chiral symmetry breaking, while for $\frac{3}{2}(N_c-2)<N_F<3 (N_c-2)$ it flows to a (super)-conformal fixed point. This suggests that the free magnetic and abelian Coulomb phases are rather special to supersymmetry, and are lifted as soon as SUSY is broken. We have also paid special attention to the loop operators of the theory that can be used as proper order parameters. In the most interesting case of $Spin(N_c)$ (in which case the electric Wilson loop in the spinor can not be screened by the dynamical matter fields) we indeed find an area law behavior corresponding to true confinement for all $N_F\leq\frac{3}{2}(N_c-2)$, while the $SU(N_F)$ global symmetry is broken to $SO(N_F)$. The dynamics leading to confinement is monopole condensation. This is most clearly seen for the $N_F=N_c-2$ special case, where massless monopoles (and dyons) indeed appear at special points in the moduli space. With AMSB, we indeed find a non-vanishing monopole condensate in accordance with the original conjecture by Mandelstam and 't~Hooft. By considering mass deformations to the $N_F=N_c-2$ case, we have numerically verified that the monopole condensate persists for all $N_F\leq N_c-2$. For the $N_c-2<N_F \leq \frac{3}{2}(N_c-2)$, the AMSB vacuum is obtained when the quarks of the dual theory are integrated out and the dual theory becomes pure YM, for which we already established monopole condensation as the special case $N_F=0$.

\section{Anomaly Mediation}\label{sec:App}

Anomaly mediation of supersymmetry breaking (AMSB) is parameterized by a single number $m$ that explicitly breaks supersymmetry in two different ways. One is the tree-level contribution based on the superpotential
\begin{align}
	{V}_{\rm tree} &= m \left( \varphi_{i} \frac{\partial W}{\partial \varphi_{i}} - 3 W \right)
	+ c.c. \label{eq:AMSBW}
\end{align}
The other is the loop-level supersymmetry breaking effects in tri-linear couplings, scalar masses, and gaugino masses \cite{Pomarol:1999ie,Randall:1998uk},
\begin{align}\label{eq:AMSBloop}
	A_{ijk} (\mu) &= - \frac{1}{2} (\gamma_{i} + \gamma_{j} + \gamma_{k})(\mu)\, m, \\
	m_{i}^{2}(\mu) &= - \frac{1}{4} \dot{\gamma}_{i}(\mu)\, m^{2}, \\
	m_{\lambda}(\mu) &= - \frac{\beta(g^{2})}{2g^{2}}(\mu)\, m.
\end{align}
Here, $\gamma_{i} = \mu\frac{d}{d\mu} \ln Z_{i}(\mu)$, $\dot{\gamma} = \mu \frac{d}{d\mu} \gamma_{i}$, and $\beta(g^{2}) = \mu \frac{d}{d\mu} g^{2}$. When the gauge theory is asymptotically free, $m_i^2>0$ which stabilizes the theory against run-away behaviors.
Note that Eqs.~(\ref{eq:AMSBW},\ref{eq:AMSBloop}) also break the $U(1)_R$ symmetry explicitly and hence we do not need to study its anomaly matching conditions.

\bibliographystyle{utcaps_mod}
\bibliography{AS_SO}

\end{document}